\documentclass[prl,twocolumn,superscriptaddress,groupedaddress]{revtex4}  % for review and submission
\usepackage{graphicx}  % needed for figures
\usepackage{dcolumn}   % needed for some tables
\usepackage{bm}        % for math
\usepackage{amssymb}   % for math
\usepackage{amsmath}
\usepackage[utf8]{inputenc}  
\usepackage[T1]{fontenc}

\usepackage{xcolor}
\usepackage{parskip}
\usepackage{siunitx}
\usepackage{url}

\usepackage{hyperref}

\begin{document}

\title{Time-Resolved Thermal Susceptibility Mapping via Low-Temperature Scanning Laser Microscopy}

\author{N. Lejeune}
\affiliation{Experimental Physics of Nanostructured Materials, Q-MAT, Department of Physics, Universit\'{e} de Li\`{e}ge, B-4000 Sart Tilman, Belgium}

\author{S. Van der Heyde}
\affiliation{Experimental Physics of Nanostructured Materials, Q-MAT, Department of Physics, Universit\'{e} de Li\`{e}ge, B-4000 Sart Tilman, Belgium}

\author{A. V. Silhanek}
\affiliation{Experimental Physics of Nanostructured Materials, Q-MAT, Department of Physics, Universit\'{e} de Li\`{e}ge, B-4000 Sart Tilman, Belgium}

\date{\today}

\begin{abstract}
We present a multiharmonic lock-in detection approach that utilizes the inverse Fast Fourier Transform to reconstruct the time evolution of thermal susceptibility with high spatial resolution. The method is implemented on a custom-built, modular scanning laser microscope designed for operation in low working-distance optical cryostats and thoroughly calibrated for spatial accuracy. A demonstrative case study using a superconducting resonator highlights the capability of this technique to generate thermal images on time scales significantly shorter than the intrinsic scanning speed, thus enabling dynamic thermal characterization with high spatial fidelity. The proposed technique of low-temperature time-resolved scanning laser microscopy offers unique opportunities to explore superconducting devices, 2D materials, hybrid planar structures, and other low-dimensional systems. 
\end{abstract}
\maketitle

\section{Introduction}

% Describe the general principle of a SLM
Scanning laser microscopy (SLM) is a technique that raster-scans the surface of a sample under test (SUT) pixel by pixel using a focused laser beam as the optical probe. The interaction between light and the SUT locally alters the properties of the sample (or the probe) and gives rise to a position-dependent photoresponse, Pr$(x,y)$.
\\
\\
This is a technique which has reached maturity, is commercially available, and is regularly applied across a large range of disciplines, including biology \cite{pawley2006handbook,Paddock2000}, engineering \cite{Nikoonahad}, and material science \cite{Tata1998}. Several books \cite{WilsonSheppard1984ScanningOptical} and review articles \cite{Minsky1988LaserScanningMemoir,Alford,Ciovati,Yann} have examined the fundamentals of this technique, its practical realization, and proposed robust instrument design.
\\
\\
Depending on the chosen photoresponse signal, a diversity of imaging modes can be implemented. These modes can be selected based on the SUT and the specific sample properties to be probed. The most common imaging mode is reflectometry, in which the photoresponse Pr$(x,y)$ is modulated by variations in the local refractive index and surface orientation.
\\
\\
SLM has been particularly useful in the study of superconducting films, for which the integration of the setup in an optical cryostat is essential \cite{Clem-1980,divin1991laser,Holm,sivakov1996spatially,Kittel,Zhuravel,Murakami,Klein,Lange,Bezard}. In the case of a current-carrying superconducting film, it is convenient to choose as the response signal the change in the voltage drop across the sample that arises from the modulation in resistance at the illuminated spot. The primary mechanism governing the photoresponse is the resistance increase due to lattice heating induced by low-energy phonons. Additionally, the Cooper-pair-breaking process induced by photons, leading to the generation of high-energy quasiparticles, may need to be considered \cite{testardi1971destruction}. Several alternative imaging modes revealing the superconducting properties of the SUT, other than optical and voltage contrast, are described in Ref. \cite{Zhuravel}. 

Unlike wide-field imaging, scanning laser microscopy suffers from inherently low-speed acquisition, typically in the range of 10$^3$ px/s, which translates into a few minutes for a 512 $\times$ 512 px$^2$ image. Therefore, this technique is well-suited to investigating the quasi-static properties of the sample but offers limited access to short-time-scale processes. Achieving high spatial resolution while capturing dynamic effects still remains a challenging task.  
\\
\\
A possible strategy to overcome this limitation is the so-called lazy fisherman method \cite{kohen2005superconducting,timmermans2014dynamic,kramer2010direct,embon2015probing}, in which the probe remains stationary at a single point while the system is driven by a periodic signal (electric, magnetic, elastic, etc.). Using the AC reference signal to set the time counter and assuming that the system response is also periodic, one can combine the periodic signal picked up at each pixel to construct a two-dimensional map revealing the oscillatory response of the system over time. If the excitation is not periodic, the collected signal represents a time-average, as illustrated for instance in Ref. \cite{silhanek2010formation,embon2017imaging,nishimura2025visualization}.

An important extension for systems driven by periodic excitations was introduced by Borgani et al. \cite{borgani2019fast}, permitting to achieve sub-millisecond time-resolved conductive atomic force microscopy measurements. Applying this method to SLM requires the sample to be excited by an amplitude-modulated laser beam at a fundamental frequency $f_1$, while a large number of harmonic components of the photoresponse at integer multiples $nf_1$ should be simultaneously acquired using lock-in detection. The frequency-domain data thus obtained can then be transformed into the corresponding time-domain response through an inverse Fourier transform. A somewhat similar approach has been used to reconstruct magnetization loops from magnetic susceptibility harmonic data \cite{Silhanek-AC,Takashi}. In the present work, we apply this concept using hundreds of harmonics to reconstruct thermal-susceptibility images of a high-$T_c$ superconducting resonator with a few-µs time resolution. We demonstrate that the proposed technique constitutes a powerful, broadly applicable tool for resolving rapid thermal phenomena across a wide range of condensed‑matter systems.  

\section{Experimental}
\subsection{Apparatus}

\begin{figure*}
    \centering
    \includegraphics[width=\linewidth]{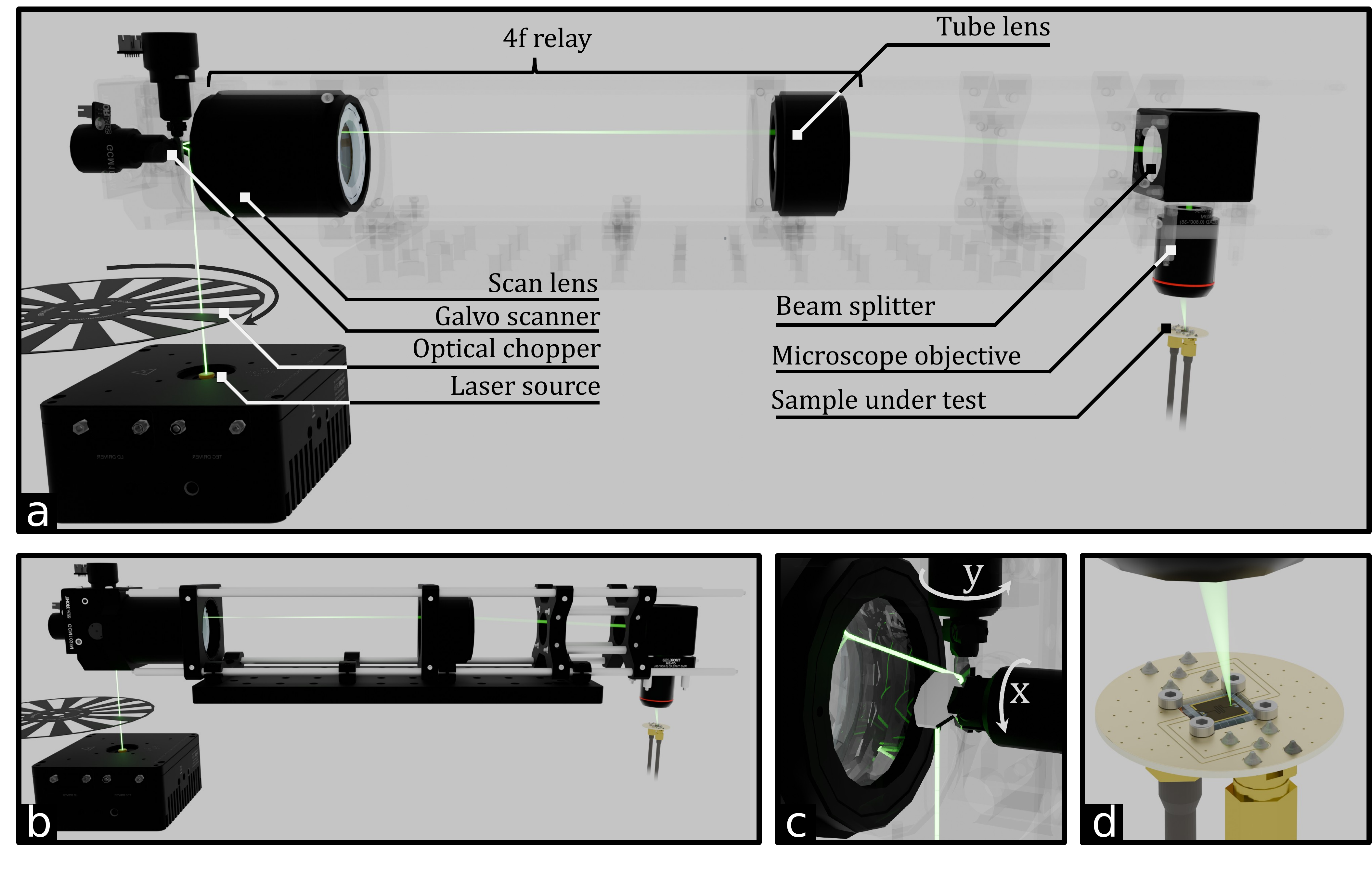}
    \caption{(a) Schematic view of the optical elements of the scanning laser microscope. The collimated laser beam passes through a mechanical chopper and is then steered using a dual-axis galvo scanner. The beam then passes through a 4f-relay system and is reflected towards the objective pupil using a 50-50 beam splitter. The light reflected from the sample can be coupled to a photodiode (not shown) using a converging lens placed above the beam splitter. (b) Schematic view of the entire mechanical setup. It is assembled as a single module that can be adapted to fit a wide range of measurement platforms. (c) Close-up view of the 2-axis scanner whose center is placed at the focal point of the scan lens. (d) Sample under test, in this example a superconducting resonator, electrically connected to the electrical instruments used for thermal susceptibility measurements.}
    \label{fig:fig1}
\end{figure*}

The central architecture of the scanning laser microscope is illustrated in Fig. \ref{fig:fig1}(a) (optical elements) and Fig. \ref{fig:fig1}(b) (integrated assembly). The system comprises a collimated laser source coupled to a dual-axis galvanometer scanner, which deflects the beam toward a scan lens (see Fig. \ref{fig:fig1}(c) for a detailed view). The mirror assembly and the scan lens are positioned such that the mean pivot point of the scanners coincides with the back-focal plane of the scan lens. Consequently, the laser beam emerges from the scan lens parallel to the primary optical axis of the microscope. The beam subsequently passes through a tube lens, positioned so that its back focal point coincides with the image point of the scan lens, thereby establishing a 4$f$ relay system. Eventually, the laser beam is reflected by a 50/50 beam splitter towards a microscope objective placed at the converging point of the tube lens, in such a way that the laser enters the pupil of the objective with a pivot angle that is linearly converted to a lateral displacement of the focused beam on the sample under test (see Fig \ref{fig:fig1}(d)). 

The microscope parts are assembled in a single rigid structure on an optical breadboard and therefore can be readily attached to an existing experimental setup. By chopping the incoming laser beam at a relatively low frequency (0.1-1 kHz), we are able to measure the local thermal susceptibility of the sample under localized heating, using a lock-in amplifier (Stanford Research System, model SR830). The chopper is placed right after the laser beam source to ensure that the phase of the chopped light beam is consistent throughout the entire scanning area.

To facilitate the positioning of the sample and therefore the interpretation of the electrical measurements, a converging lens collects the light reflected off the sample and passing through the beam splitter and focuses it on a reverse-biased photodiode. This approach offers two operating modes: without the chopper, the reflected light is measured using a conventional transimpedance amplifier and voltmeter triggered at each sampling point, whereas with the laser chopped, the reflectivity can still be measured using a lock-in amplifier. This way, classical optical and electrical response images can be acquired concomitantly using two different lock-in amplifiers, each triggered by the same signal. Thus, the electrically detected susceptibility to light-induced heating can be straightforwardly superimposed on top of the raster-scanned optical image without the need for alignment, as they are acquired simultaneously.

Caution should be exercised when quantitatively interpreting the thermal susceptibility maps. In the general case, in which the sample under test has regions of significantly different reflectivity, one must take into account the fact that the heat periodically introduced at each pixel spot is not uniform all over the sample, but depends on the local reflectivity and material properties. Measuring reflected light can be instrumental in estimating the amount of light absorbed and in normalizing the detected electric response.

\subsection{Scan calibration and reflective spot size}

\begin{figure}
    \centering
    \includegraphics[width=\linewidth]{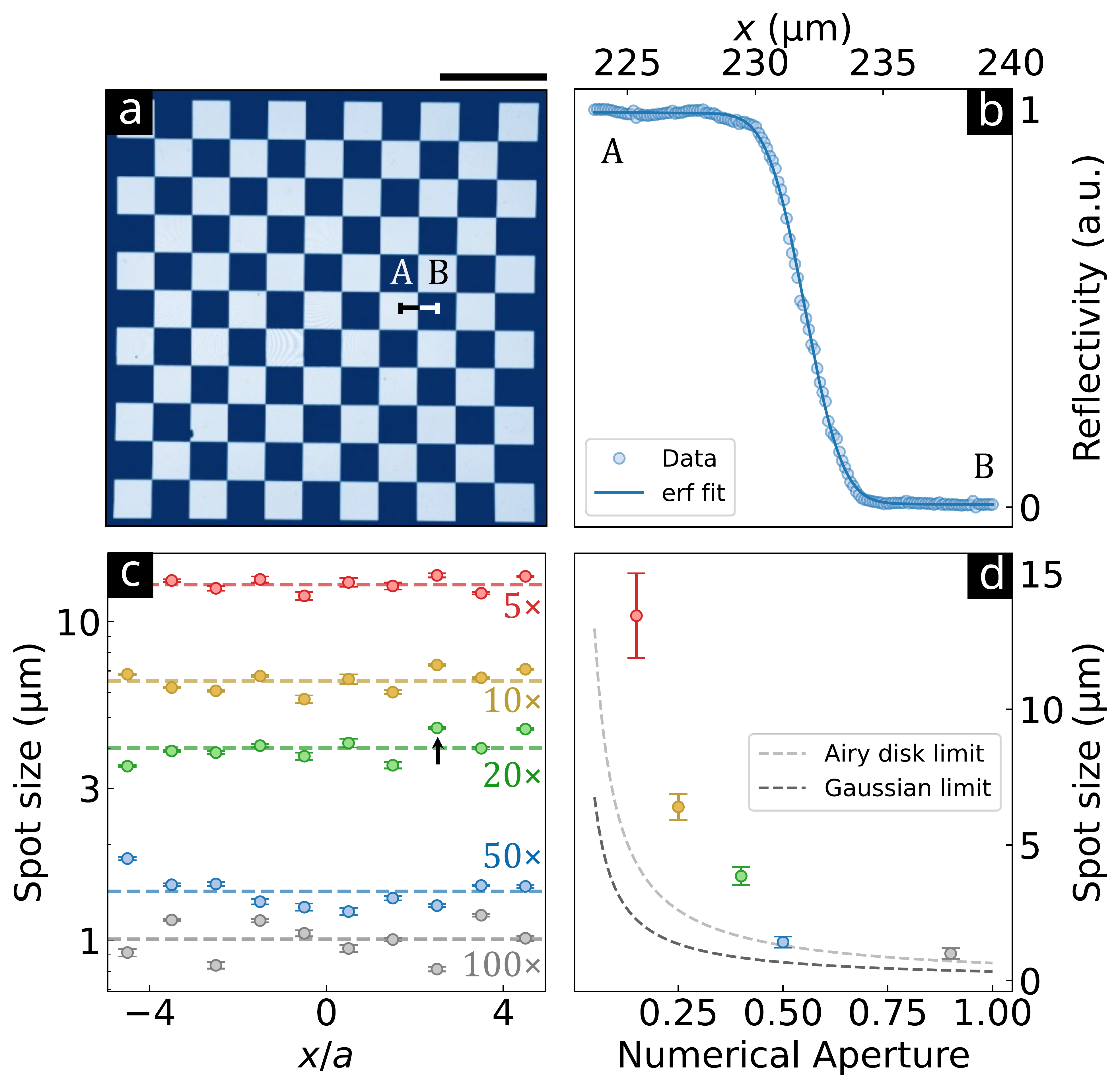}
    \caption{(a) A representative 501$\times$501 px$^2$ SLM image acquired using the sample reflectivity as photoresponse. This image was obtained using a 20x objective and a photodiode as detector. The scale bar in the upper right corner corresponds to 250 µm. (b) Line scan along the segment A-B indicated in panel (a) spanning across the transition between the substrate and the reflective Al square. The solid line corresponds to the fit of Eq. \ref{eq:erf} and is used to estimate the $1/e^2$ beam diameter (spot size). (c) Measured spot size at various $x$-axis scan positions, normalized by the size of a single square cell of the chessboard ($272.8$, $272.8$, $90.9$, $22.7$ and $11.4$ µm, for 5x, 10x, 20x, 50x and 100x objectives, respectively). The mean values are indicated with dashed lines, and the spot size associated with the fit in panel (b) is highlighted by a black arrow. (d) Mean beam diameter values for the different tested objectives. The light gray dashed line corresponds to the Airy disk limit given by $\mathrm{spot\;size}  = \frac{1.22 \:\lambda}{\mathrm{NA}}$. Whereas the dark gray dashed line corresponds to the Gaussian beam limit given by $\mathrm{spot\;size}  = \frac{0.635 \:\lambda}{\mathrm{NA}}$.} 
    \label{fig:2}
\end{figure}

The X/Y voltage-to-distance conversion factor, the distortion amplitude, and the spot size are objective-dependent. Therefore, it is necessary to carry out a calibration procedure for each magnification used. To do so, calibration chessboards of different lateral sizes (each having $11\times 11$ squares) made of reflective 50-nm-thick Al were patterned on top of Si/SiO$_2$\,(300 nm) substrates. Fig. \ref{fig:2}(a) shows one of the chessboard patterns as acquired by the SLM with a 20x objective using 501 $\times$ 501 px$^2$ resolution. In this case, the photoresponse is simply the material's reflectivity. An example of a line scan Pr$(x,y_0)$ along the segment A-B is shown in panel (b). This position-dependent photoresponse is fitted using:

\begin{equation}
    \mathrm{Pr}(x,y_0)= A\ \mathrm{erf}\left(\frac{x-x_0}{\sqrt{2}\sigma}\right)+Bx+C,
\label{eq:erf}
\end{equation}

\noindent with $\mathrm{erf}(x)$ the error function, $x_0$ the position of the border of the corresponding Al square, $\sigma$ the standard deviation, and $A$, $B$, $C$ fitting constants. Fig. \ref{fig:2}(c) shows the $1/e^2$ spot size of the focused beam, defined as 4$\sigma$, for each border of the square cases of the board. This figure shows that the beam exhibits little distortion across the entire field of view for magnifications below 50x, and a tendency to increase in size for 50x and 100x objectives near the field-of-view boundaries. We have also investigated pillow-shaped field distortion (pincushion distortion). This distortion is caused by the mirror arrangement and results from magnification increasing with distance from the optical axis, thereby causing straight lines to bow inward. By using a hyperbolic fit, we found that these distortions have a maximum amplitude of 6 µm for a line scan of 1 mm, and can be easily corrected if needed, although they represent a negligible source of error which can typically be ignored. In Fig. \ref{fig:2}(d), the dependence of the spot size on the numerical aperture of the objectives is shown. The gray dashed line shows the Airy disk diameter $d=1.22 \lambda/\mathrm{NA}$ (where NA is the numerical aperture of the used objective) expected from the diffraction limit for a laser source of wavelength $\lambda=532$ nm. A more rigorous derivation accounting for the Gaussian character of the laser beam yields a smaller diffraction limit \cite{Urey}, given by $d=0.635 \lambda/\mathrm{NA}$ and represented by the dashed black line.

\subsection{Thermal spot size}

\begin{figure}
    \centering
    \includegraphics[width=0.98\linewidth]{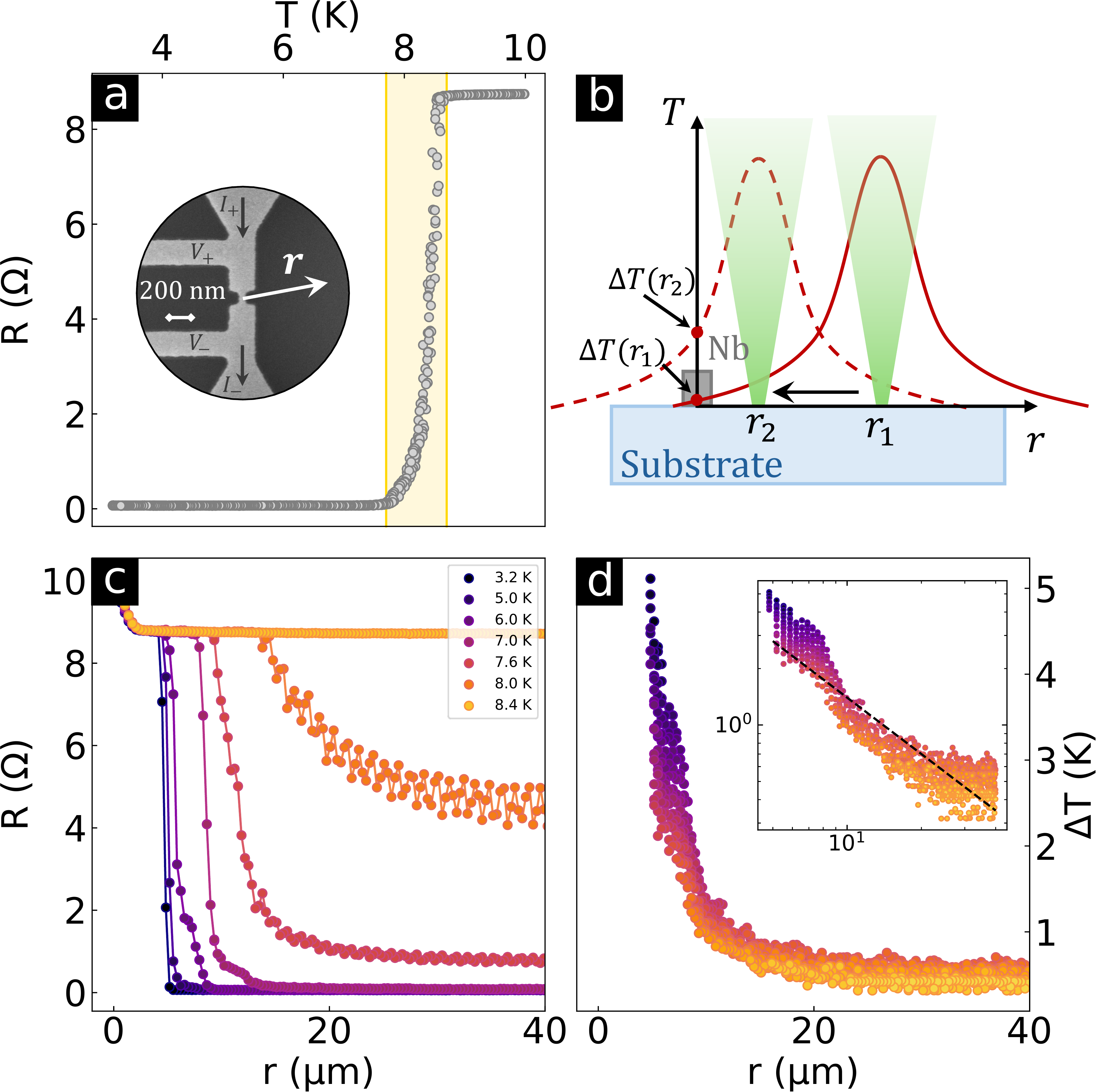}
    \caption{(a) Resistance of a Nb junction thermometer as a function of temperature. The range over which the resistance-to-temperature conversion is exploitable is shaded in yellow. The inset shows a scanning electron microscope image of the probing device: the current is injected through the top and bottom terminals, and the voltage is measured across the left-hand lines. The distance $\boldsymbol{r}$ defines the reference used in panels (b-d). (b) Schematic of the temperature profile induced by the absorbed laser light. The Nb junction measures the temperature rise as the laser is scanned across the surface. (c) Resistance as a function of the distance between the center of the laser spot and the center of the junction, for various base temperatures. The increase in resistance occurs when the laser reaches a distance where the temperature profile drives the junction through its superconducting-to-normal transition at $T_c$. The additional increase in resistance at distances of a few microns corresponds to the laser directly illuminating the junction and heating it well above $T_c$. (d) Reconstruction of the thermal profile, with each curve in (c) contributing to one region of the profile, as indicated by the color coding. The inset shows the same data on a log--log scale, where the dashed line corresponds to $\Delta T\propto r^{-1}$.}
    \label{fig:fig3}
\end{figure}

In order to characterize the temperature profile generated by light absorption, a 50-nm-wide Nb bridge in a four-wire configuration, fabricated on top of a Si/SiO$_2$ substrate (see inset of Fig.~\ref{fig:fig3}(a)), is used as a thermometer. The resistance of the superconducting junction as a function of temperature is shown in Fig.~\ref{fig:fig3}(a). The laser is scanned across the surface while the temperature at the junction is monitored through its resistance. Tracking this temperature as a function of the laser spot position amounts to measuring the temperature profile, as represented schematically in Fig.~\ref{fig:fig3}(b). This approach permits resolving the temperature profile as long as the laser spot lies in the region where it induces a temperature increase $\Delta T$ at the junction, such that $T_\mathrm{base}+\Delta T\approx T_c$, where $T_\mathrm{base}$ is the temperature of the cold finger to which the back side of the sample is thermally anchored. When the cold finger is kept at low temperature, $\Delta T$ must be large and the profile is measured over small radial distances $r$ from the junction (see inset of Fig.~\ref{fig:fig3}(a)). Conversely, when the cold finger is held at a temperature close to but below $T_c$, $\Delta T$ can be small and the profile is resolved over large distances.

%$2.5$\,mW
Following this method, $R(r)$ curves are acquired at various fixed $T_\mathrm{base}$, as shown in Fig.~\ref{fig:fig3}(c). When the laser irradiation approaches the junction, a sudden resistance increase is observed once the junction temperature reaches the vicinity of $T_c \approx 8.5$\,K. From these curves, the temperature profile $\Delta T(r)$ can be reconstructed, as depicted in Fig.~\ref{fig:fig3}(d). The temperature profile approaches a $1/r$ law (see inset of panel (d)), as expected for a steady point source~\cite{Carslaw_1959}. The proposed method is less sensitive to resolving the temperature in the core of the optical spot because, as visible in Fig.~\ref{fig:fig3}(c), even at the lowest cold finger temperature, the junction reaches the normal state at a distance of approximately 5\,\textmu m. Yet, a clear increase in the normal state resistance is observed when the laser is close to the junction, permitting us to estimate a substantial increase in temperature of about 50\,K at the center of the illumination spot. This is further enhanced by the fact that the heat is deposited directly into the Nb, rather than in the substrate, from which it would otherwise have to diffuse through the substrate--device interface with a non-zero thermal resistance. This effect is balanced by the fact that the Nb junction reflects more light than the substrate, which in turn reduces the thermal absorption. The above estimations correspond to a laser power of $2.5$\,mW and depend linearly on the incident power~\cite{Carslaw_1959}. In summary, the laser-induced heating is intense and varies rapidly at short distances, then extends farther from the device following a weaker spatial dependence. This far-range characteristic distance is substrate-dependent through the thermal transport characteristics of the substrate.

\section{Data acquisition and FFT post-processing}

For the sake of clarity, it is convenient to start by reminding the basic principle of a lock-in amplifier (LIA). This instrument contains an internal oscillator generating two $90$\textdegree\ phase-shifted references, $R_{x,n}(t)$ and $R_{y,n}(t)$, which can be phase-locked to an external TTL triggering signal. If no phase-shift relative to the sample excitation is configured (auto-phase), and assuming the trigger event defined by the LIA occurs at $t=t_0$, 

\begin{align}
    R_{x,n}(t)&=\sin(2\pi nf_1(t-t_0))\\
    R_{y,n}(t)&=-\cos(2\pi nf_1(t-t_0)),
\end{align}

\noindent where $f_1$ is the frequency of the reference signal and $n$ is an integer multiple of the fundamental frequency $f_1$ (i.e. the $n$-th harmonic). For the particular case of the Stanford Research SR830 LIA used in this work, $n$ is restricted to values such that both conditions $n<19999$ and $nf_1<102$ kHz are satisfied. 
The measured voltage signal $V_{in}(t)$, which can be either single-ended or differential, DC or AC coupled, is multiplied by both reference signals and passed through a low-pass filter of time constant $\tau$, which is user-defined. This analogical processing is mathematically equivalent to 

\begin{equation}
    V_{x,n}(\mathcal{T})=\int_{\mathcal{T}-\tau}^{\mathcal{T}}V_{in}(t)R_{x,n}(t)\,\mathrm{dt},
\end{equation}

\noindent and similarly for $V_{y,n}$, using $R_{y,n}$. It is worth noting that the technique is restricted to perfectly periodic Pr$(x,y)$, and that time-resolved reconstruction will not work for samples with hysteretic behavior. In the case of periodic signals and for $\tau\gg1/(nf_1)$, the output of the LIA corresponds to the Fourier coefficients of the input signal. Since the LIA uses a sine reference to measure the in-phase signal, the complex Fourier coefficients needed to compute the inverse transform are given by

\begin{equation}
    C_n=-V_{y,n}+iV_{x,n},
\end{equation}

\noindent where $C_n$ is the Fourier coefficient of the $n^{th}$ harmonic and $i^2=-1$. 

Collecting a sufficient number of harmonics allows for accurate computation of the inverse Fourier transform pixel-by-pixel. The result is a time evolution of the local thermal susceptibility of the SUT within a time window given by $\Delta t=1/f_1$. For $N$ collected harmonics, $2N$ frames spaced in time by $\delta t=(2Nf_1)^{-1}=\Delta t/2N$ are obtained. Because the reconstructed signal must be purely real, the imposed Hermiticity adds the factor $2$, effectively doubling the temporal resolution. Therefore, irrespective of the chopping frequency, the maximum temporal resolution achievable is defined by the maximum harmonic the LIA can measure, which in our case corresponds to $\delta t_{min}\approx 1/204\textrm{ kHz}\approx4.9 \textrm{ µs}$. In principle, a similar procedure could be achieved using an oscilloscope, but with much poorer resolution. 

It is important to note that the time-domain reconstruction should not be misinterpreted as a direct visualization of the local thermal relaxation following localized heating. Although the reciprocal frames are evenly spaced in time, the mapped quantity represents the {\it global} dynamic susceptibility Pr$(x,y)$ to the \textit{locally} applied stimulus. The temporal evolution registered at an individual pixel serves as a proxy for the thermal diffusion rate toward the highly sensitive regions of the device, parameterized by the phase shift relative to the periodic excitation. Ultimately, this represents a mathematical transformation from the frequency domain rather than a strict spatiotemporal thermograph.

\begin{figure}
    \centering
    \includegraphics[width=\linewidth]{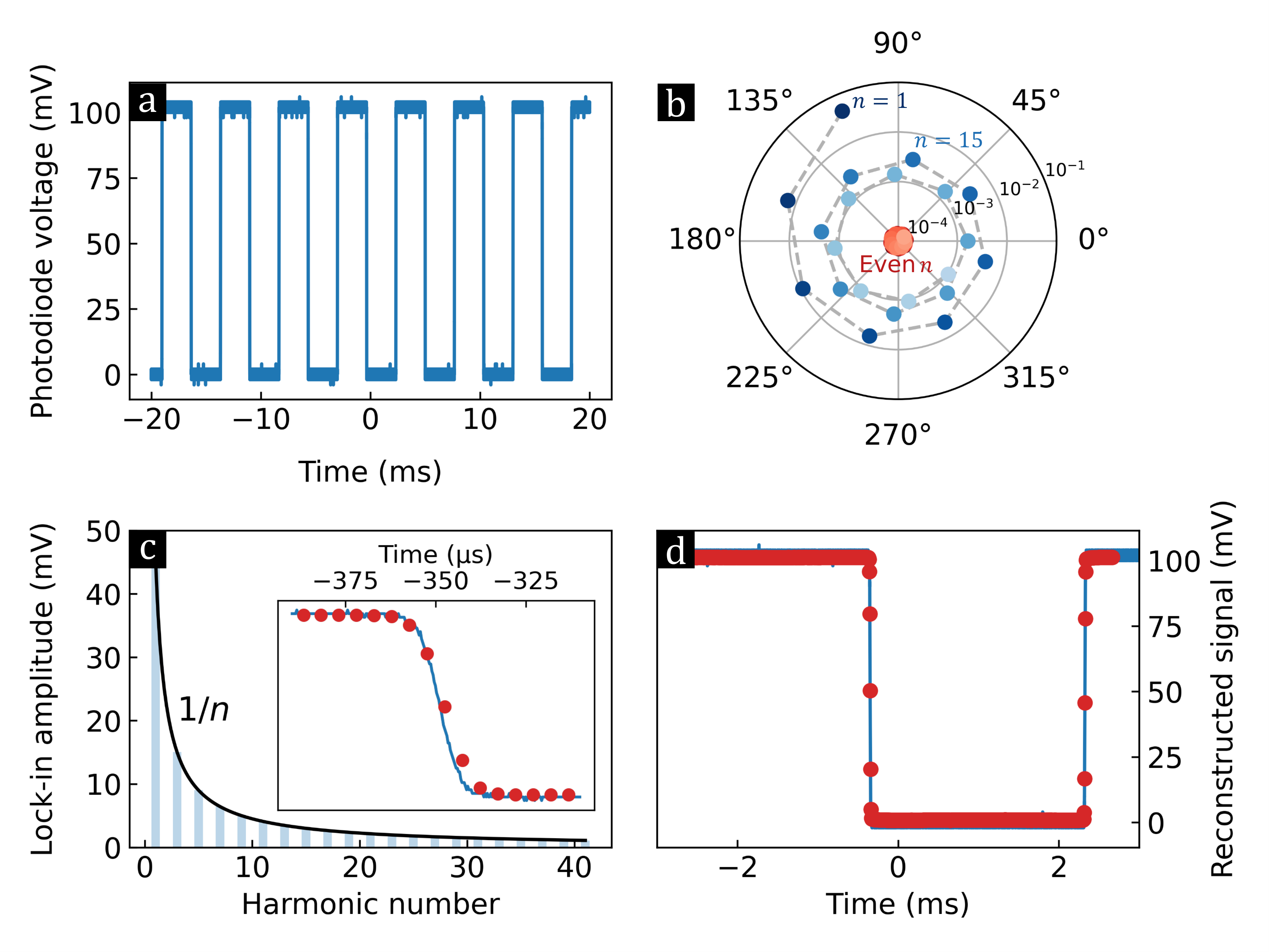}
    \caption{Validation and proof-of-principle. (a) A laser is chopped at a fixed frequency $f=188.27$ Hz, and the light reflected in a mirror is detected using a photodiode. The signal is periodic and can therefore be reconstructed using the inverse Fourier transform method. (b) Coefficient measured using a lock-in amplifier for the first 40 harmonics. The first harmonic $n=1$ is the strongest. The amplitude decreases inversely proportional to the odd harmonic number, whereas the even harmonics are significantly smaller. The information carried by the phase corresponds to the synchronization of the detected signal to the reference TTL signal. (c) Amplitude of the measured Fourier coefficients. (d) Reconstructed signal (red) plotted on top of the signal of panel (a). The inset in panel (c) shows a zoom on the downward transition in the photodiode voltage and emphasizes the excellent fidelity of the reconstruction to the original signal.}
    \label{fig:fig4}
\end{figure}

In order to validate the technique, we first apply it to reconstruct the temporal evolution of light reflected from a mirror. In this case, the detector is a photodiode, which should pick up a square wave photoresponse resulting from a laser source modulated by a mechanical chopper.

The time-domain voltage through the reverse-biased photodiode measured using the oscilloscope is shown in Fig. \ref{fig:fig4}(a) for a chopping frequency $f_1=188.27$ Hz. Next, in order to obtain the frequency-domain response, the photodiode's output is connected to the LIA's input, and the values of $V_{x,n}$ and $V_{y,n}$ are collected for all accessible harmonics. The sensitivity of the instrument is progressively adjusted as the amplitudes of the Fourier coefficients decrease. The so-obtained Fourier coefficients, plotted in a polar graph with amplitude and phase, are shown in Fig. \ref{fig:fig4}(b). Because the inverse FFT takes as input the coefficients in both the negative and positive frequency domains, the complex Fourier coefficients are extended in a Hermitian manner, thus ensuring that the temporal reconstruction is a purely real function. Notably, the DC coefficient at $f=0$ Hz cannot be measured by the LIA, but must be measured using a voltmeter to fully reconstruct the inverse FFT. However, the effect of this DC component is simply to shift the inverse FFT vertically and therefore does not contain information about the temporal dynamics of the response. Unless explicitly mentioned, we arbitrarily set it to zero, such that the reconstructed temporal evolution will be centered around zero. Fig. \ref{fig:fig4}(c) shows the amplitude of the Fourier coefficient. As expected for a square wave, the amplitude decreases inversely proportional
to the odd harmonic number, whereas even harmonics are negligibly small. Fig. \ref{fig:fig4}(d) shows the reconstructed signal (red dots) obtained from the inverse FFT of the frequency-domain measurements plotted on
top of the time-domain signal of panel (a). The inset in panel (c) shows a zoom on the downward transition in the photodiode voltage, emphasizing the excellent fidelity of the reconstruction to
the original signal.

\section{Case study: Superconducting resonator}

\begin{figure*}
    \centering
    \includegraphics[width=\linewidth]{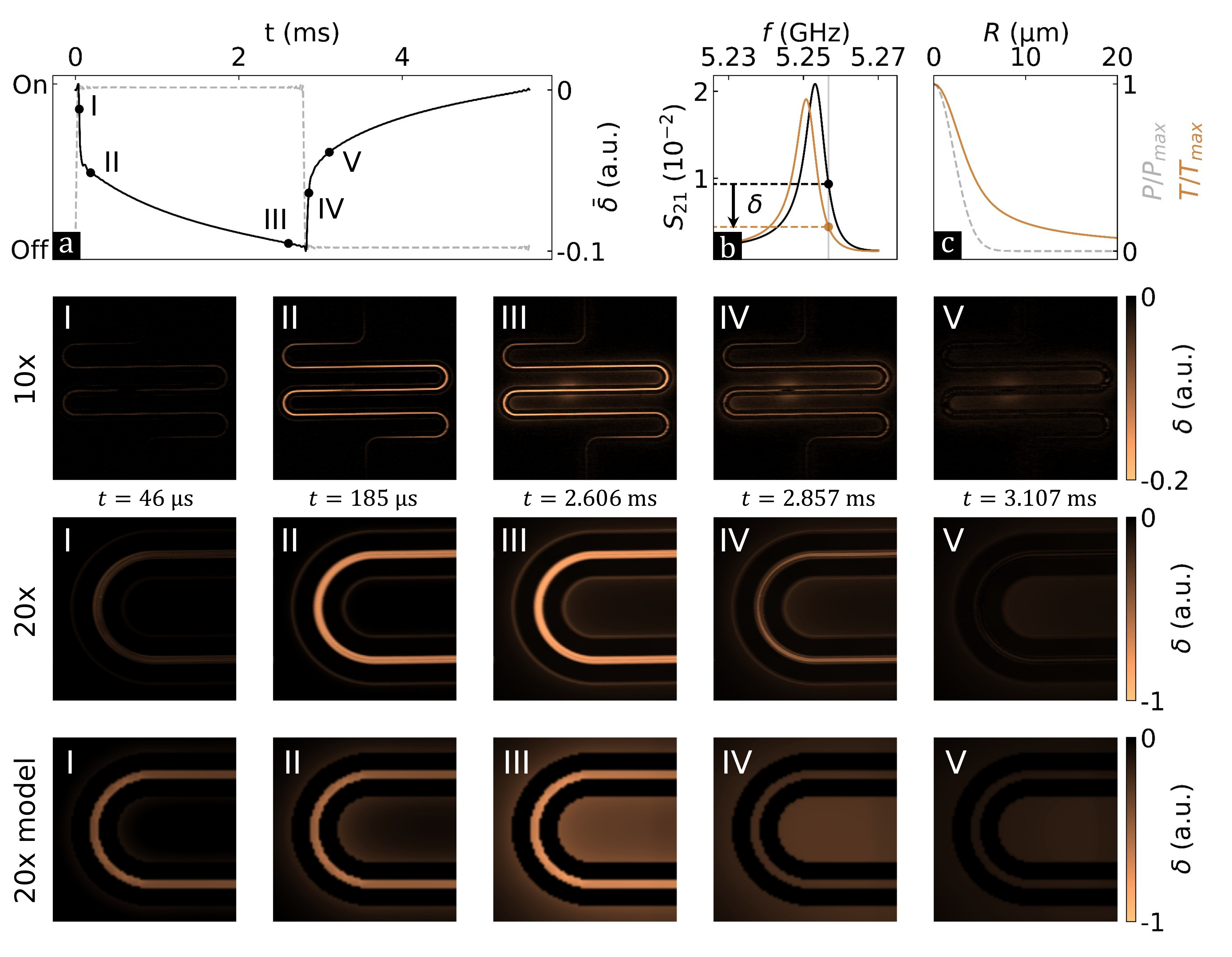}
    \caption{(a) Time-resolved reconstruction of the average response of the superconducting resonator over one chopper period. The signal captured by the photodiode, corresponding to the on/off state of the laser, is plotted in a dashed line. (b) Graphical representation of the probed photoresponse $\delta$, corresponding to the change in RF power transmitted through the resonator at fixed frequency and which depends on the temperature distribution in the sample. The black (orange) curve illustrates the case where the laser is away (near) from (to) a sensitive region. (c) Optical (dashed gray) and thermal (yellow) spot profiles. The thermal profile is obtained from simulation and is used as the kernel for the convolution model of the bottom row. Snapshots at five different instants, labeled I-V and indicated in panel (a), are shown in the bottom three rows. The first row corresponds to images of the photoresponse of the entire sample using a $10\times$ objective, whereas the second row shows images collected using a $20 \times$ objective, zooming on the bottom left half-turn of the meander. The bottom row shows the result of the time-dependent convolution model developed in the text.}
    \label{fig:fig5}
\end{figure*}

The time-resolved reconstruction technique was implemented on optimally doped $\text{YBa}_2\text{Cu}_3\text{O}_{7-\delta}$ (YBCO) half-wavelength superconducting resonators. Details regarding the fabrication process are provided in Ref. \cite{Lejeune2026}. The specimen was mounted within a Montana Instruments optical cryostat. One port of the resonator was coupled to a Vector Network Analyzer operating in continuous wave mode, while the output port was connected to a $15$~dB RF amplifier followed by a Krytar 203B RF diode. The diode rectifies the RF signal into a DC voltage proportional to the transmitted power. This voltage was subsequently processed by a lock-in amplifier phase-locked to the reference frequency of the scanning laser microscope optical chopper. To characterize the device, the scattering parameter $S_{21}(f)$ was initially recorded to identify the fundamental resonance frequency. Fig. \ref{fig:fig5}(b) displays the resonance peaks obtained at $T = 30$~K (black) and $T = 32$~K (yellow). Note that the resonance frequency shifts to lower frequencies with increasing temperature, a behavior attributed to the enhancement of the material kinetic inductance. This redshift leads to a measurable change in DC output $\delta$, which is used as photoresponse. This preliminary characterization facilitates the selection of an optimal probe frequency for the LIA. During SLM operation, the modulated laser beam induces a periodic shift in the resonance peak, as illustrated in Fig. \ref{fig:fig5}(b), thereby modulating the rectified DC output voltage at the chopper frequency. In the perturbative regime, wherein the laser induces small heating of the device, the modulation of the transmitted power remains directly proportional to the shift in the resonance frequency.

SLM images were acquired using a chopping frequency of $180$~Hz. To achieve a comprehensive time-resolved reconstruction, $300$ harmonics were recorded simultaneously by employing three LIAs in parallel, with each instrument devoted to a specific harmonic range: $1$–$100$, $101$–$200$, and $201$–$300$. The high input impedance of the SR830 ($10$~M$\Omega$) facilitates the parallel coupling of multiple instruments without inducing signal distortion or loading effects. Consequently, the total acquisition time is reduced by a factor of approximately $\mathcal{N}$, where $\mathcal{N}$ represents the number of instruments utilized in parallel. To ensure precise temporal synchronization between the reconstructed frequency response and the optical excitation, the signal from the photodiode was monitored using a fourth LIA restricted to odd harmonics. This configuration provides sufficient resolution to accurately determine the onset of the laser-induced transition.

The spatially averaged microwave response of the resonator is presented in Fig. \ref{fig:fig5}(a). Upon laser excitation, an initial rapid response is observed within the first $100$\,µs. The relative shift $\delta$ is negative, as the continuous-wave probing frequency is situated on the high-frequency slope of the resonance; consequently, the thermally induced redshift of the resonance peak results in a decrease in transmission (see Fig. \ref{fig:fig5}(b)). Following this initial onset, $\delta$ continues to decrease at a significantly lower rate.

To elucidate this two-step behavior, we examine five temporal snapshots showing $\delta(x,y)$ captured during the periodic stimulus. A more detailed full animation is available as Supplementary Material. These images represent time-resolved maps of the local photothermal susceptibility. Immediately following the onset of irradiation (I), the response is localized strictly to the meander. Since the central conductor carries the highest current density (proportional to the square root of the quality factor), it exhibits the highest sensitivity and yields the fastest response time. This characteristic timescale scales as $a^2/D$, where $a$ denotes the beam waist and $D$ is the thermal diffusivity~\cite{Carslaw_1959}. 

After $185$~$\mu$s (II), the susceptibility of the meander reaches saturation, while the edges of the ground plane, characterized by lower current densities, begin to emerge. Indeed, the edge currents in the ground plane, next to the excitation line or the resonators, are about a factor $1 + 2g/w \sim 2$ smaller than in the central line \cite{Culbertson}, where $g$ is the gap between the central track and ground plane, and $w$ is the width of the central track. Approaching the end of the heating phase (III), the central conductor remains saturated, whereas the ground plane exhibits a pronounced susceptibility. This suggests that over these extended timescales, heat deposited in the ground plane diffuses through the substrate to the central conductor, resulting in a delayed contribution to the reconstructed signal.

This interpretation is further supported by the dynamics observed immediately following laser extinction (IV): the susceptibility of the central conductor decays rapidly, while the signal from the ground plane persists. This behavior demonstrates a clear phase delay governed by the finite thermal diffusivity of the system. Finally, the cycle concludes as heat dissipates through both the film and the underlying substrate. 

To substantiate these claims, a modeling approach analogous to that described in Ref. \cite{Lejeune2026} was implemented. This model relies on the principle that the resonance frequency of the resonator is determined by the total capacitance and inductance of the device. Because the inductance contains a kinetic contribution from the superconducting condensate, its value depends directly on the local temperature. Furthermore, the magnitude of the resonance frequency shift is proportional to the current density circulating through the regions of varying kinetic inductance. Consequently, local temperature fluctuations in areas supporting negligible current density yield no observable effect on the global response, and conversely.

This rationale motivates the introduction of a convolution product between the square of the local current density and the spatial temperature profile induced by the laser. To implement this, COMSOL Multiphysics was employed to simulate the time-dependent temperature profile generated by the laser illumination. The optical profile used as input for the simulation is shown in dashed gray in Fig.~\ref{fig:fig5}(c), whereas the resulting induced temperature at saturation is shown in yellow in the same panel. The numerical simulation models a $500\ \text{µm}$-thick (La$_{0.3}$Sr$_{0.7}$)(Al$_{0.65}$Ta$_{0.35}$)O$_3$ slab with parameters $\kappa = 1.5\ \text{W/mK}$ \cite{Langen}, $C_v = 5.33\ \text{J/kgK}$, $\rho = 6740\ \text{kg/m}^3$, and thermalized to $30\ \text{K}$ at the bottom surface. Those parameters were optimized to match the experimental response of the sample. Unfortunately, no accurate low-temperature specific heat measurement for this particular substrate was found in the literature. 

The result of this model is shown in the bottom row of Fig.~\ref{fig:fig5}. The simulated maps reproduce the main features of the experiment: the response appears first on the central conductor of the meander, the edges of the ground plane emerge only later. Next, after the laser is switched off, the ground-plane signal persists while the central conductor decays rapidly. The agreement between the experimental results and the model supports the interpretation that the two-step temporal behavior of Fig.~\ref{fig:fig5}(a) originates from the finite time required for the heat generated in the ground plane to diffuse toward the high-current-density regions.

\section{Discussion}

To conclude the analysis of the time-resolved response, we now focus on the delay that develops between the excitation and the response when the laser is positioned far from the sensitive regions (i.e. regions of high current density). Note that the harmonics of the square-wave input all propagate at different speeds and are more strongly attenuated for shorter wavelengths (\textit{i.e.}, higher harmonic excitations), as detailed in the Annex. As a consequence, this effect leads to a progressive distortion of the response, which evolves from a square wave into a sawtooth and eventually retains only the fundamental sine component as the laser is positioned away from the sensitive region. Moreover, the differing phase velocities introduce a delay (see Fig. \ref{fig:thermal_waves}(b-c)) between the maximum temperature close to the source and far from it.

\begin{figure}
    \centering
    \includegraphics[width=\linewidth]{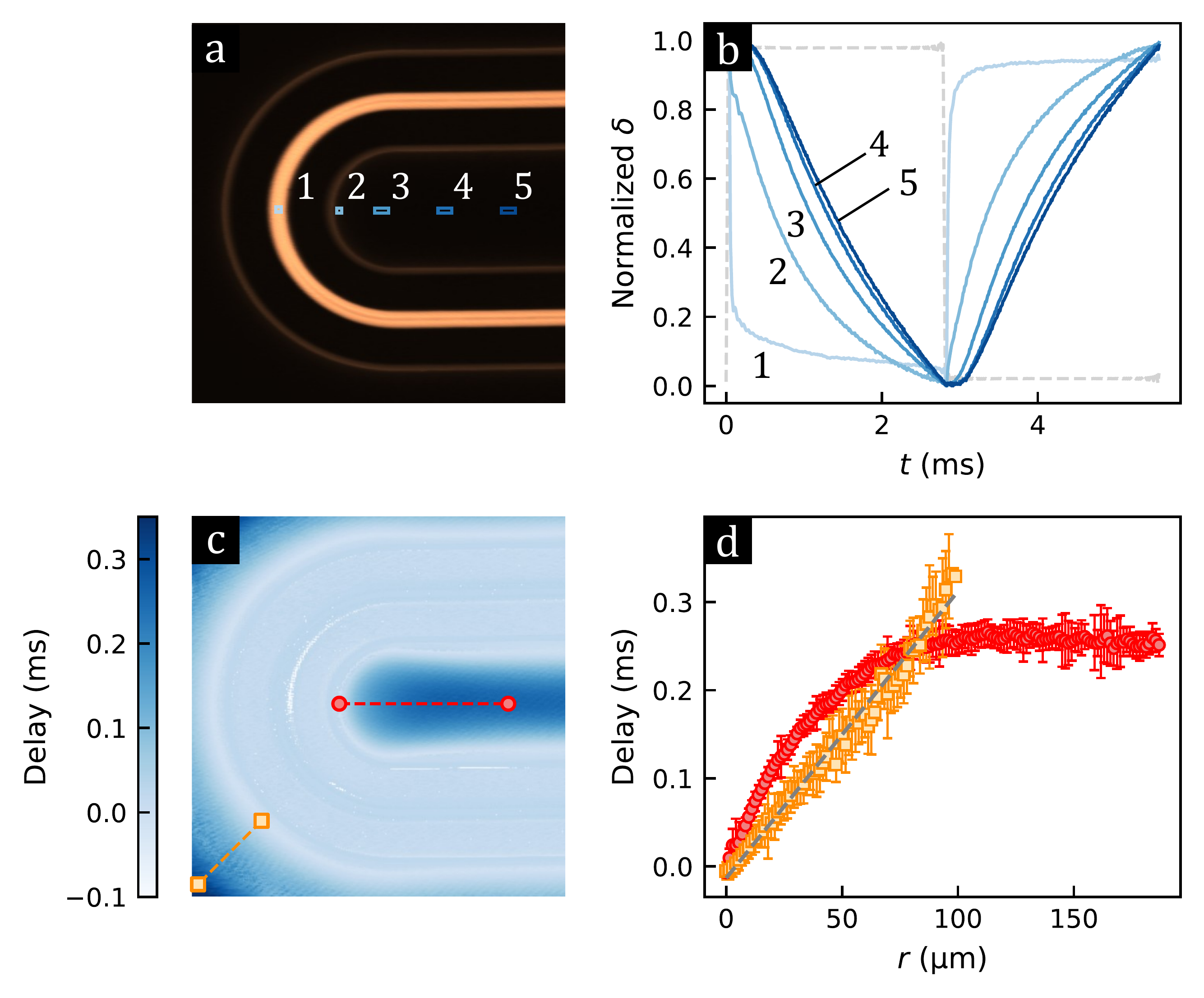}
    \caption{(a) Thermal susceptibility image of the resonator $55$~\textmu s after illumination. (b) Response of the resonator at several positions. Each probed area is marked in (a) with a rectangle. The response is normalized, and the reconstructed photodiode signal is shown as a dashed gray line. The dark-blue curves, deep in the ground plane, exhibit a delay between the laser switching and the maximum response. (c) Delay extracted pixel-by-pixel from the temporal response. The darker a region, the longer it takes for heat to propagate from that pixel to the nearest sensitive region, thereby affecting the global resonant frequency. (d) Profiles extracted from (c) along a line in the central ground plane (red) and along a line in the ground plane extending away from the resonator (orange). The red curve tends to saturate because the probed points remain within the resonator structure, between the parallel arms of the meander, whereas along the orange line the delay increases linearly with distance, as the probed point moves farther from all current-carrying regions. A propagation speed is extracted from the linear fit of the orange data, shown in gray.}
    \label{fig:fig6}
\end{figure}

Fig.~\ref{fig:fig6}(a) reproduces a snapshot of the resonator response introduced in the previous section, while its normalized local response, probed as a function of time at several selected spots, is shown in Fig.~\ref{fig:fig6}(b). The region labeled 1 is measured at the central track of the resonator meander, where the response most closely follows the excitation. The farther the probing zone lies from the sensitive regions (labeled 2-5), the more the waveform is distorted, gradually adopting a sawtooth shape.

We next turn to the delay between the switching on of the laser and the maximum of the response. Fig.~\ref{fig:fig6}(c) shows a map of this delay, extracted from the time-resolved response for each pixel. The time delay is obtained by fitting the response with a periodic saturating-exponential function with two characteristic timescales: a short one accounting for the rapid variations, and a longer one, of the order of the total period, accounting for the global response. When the laser is parked at the center of the meander, the delay is virtually zero, since the response is synchronized with the excitation. As the illuminated spot moves deeper into the ground plane, toward the center of the image, a delay of a few percent of the total cycle time builds up, visible in the dark-blue curves of Fig.~\ref{fig:fig6}(b). Cross-sectional profiles of this delay as a function of distance $r$ are shown in Fig.~\ref{fig:fig6}(d), where $r=0$ denotes the point closest to the meander half-turn. As expected, the delay increases as one moves away from the current-carrying parts, and two distinct trends emerge. The red data points show a delay that first increases and then reaches a plateau. This can be understood as follows: for values of $r$ up to the meander radius, the distance to the nearest current-carrying part grows, so the delay grows accordingly. Beyond this region, the area most strongly affected by the laser is no longer the half-turn but the straight portions of the resonator, whose distance to the cross-section line remains constant as $r$ increases, and therefore the delay saturates.

Conversely, the orange data points measure the delay radially outward from the center of the resonator. In this case, as $r$ increases, the entire current distribution of the resonator recedes from the probing point. The delay consequently increases linearly, since the current-carrying region reached first by the heat is the one located at $r=0$. Fitting the orange data with a linear law yields a heat propagation speed of $307\pm4$~\textmu m/ms. Although heat itself does not propagate like a wave \cite{Salazar_2006}, its modulation does \cite{Carslaw_1959}. The speed at which the fundamental propagates, given by $v=\sqrt{2D\omega}$ with $\omega=2\pi f_1$, is, to first approximation, the one that gives rise to the measured delay. The higher harmonics are neglected here, as they are more strongly damped and thus contribute less to the delay.

From the above expression for the speed of the fundamental harmonic, a diffusivity of $41.7\pm1.2$~mm$^2$/s is found. This value is the one used in the COMSOL simulations shown at the bottom of Fig. \ref{fig:fig5}. The minor differences between the maps obtained experimentally and modeled are justified by the overly simplified model. As noted above, all interfacial thermal effects were neglected in the modeling which considers that the LSAT substrate heats up directly.

Care must be exercised when interpreting these results. The probed quantity is the global shift of the resonance frequency of the resonator, obtained while illuminating a spot orders of magnitude smaller than the device itself. This minute spot generates a thermal profile that extends far beyond the laser waist size and takes time to diffuse. Over this diffusion length, several regions of the resonator are heated, some of which may carry a high current density. It is, however, difficult to disentangle the contribution of a strongly current-carrying region heated only slightly from that of a moderately current-carrying region heated substantially, as both may produce the same frequency shift. In other words, the notion of \textit{sensitive region} introduced earlier is more of a conceptual construct than a rigorously definable quantity.

\section{Conclusion}

In summary, we propose an experimental approach to obtain time-dependent scanning laser microscopy images by periodically illuminating each pixel, collecting multiple harmonics with a tandem of lock-in amplifiers, and then applying an inverse Fourier transform. We first validate the technique by reconstructing the square-wave photoresponse from an amplitude-modulated laser source. Subsequently, we apply this method to image a standing wave at the fundamental resonance frequency of a YBCO resonator as a function of time during laser irradiation. Regions with higher current density are more sensitive to local heating and provide a sizable photoresponse signal. The contrast variation of the images over time reflects the spreading of the thermal spot due to thermal diffusion. In principle, this technique permits estimation of the thermal parameters governing heat diffusion while revealing the more thermally susceptible regions of the sample.

\begin{acknowledgments} 
\subsection*{Acknowledgments}
The authors acknowledge financial support from Fonds de la Recherche Scientifique - FNRS under the grant CDR J.0199.25. N. L. acknowledges support from FRS-FNRS (Research Fellowships FRIA), and S.V.d.H. is a Research Fellow of the Fonds de la Recherche Scientifique - FNRS. We thank Anna Palau for allowing us to reuse the YBCO nanostructures of Ref.~\cite{Lejeune2026} in the present study.
\end{acknowledgments}

\section{Annex}\subsection{Spatially and Time-Resolved Thermal Waves Induced by Chopped Laser Heating}

This Annex provides a simplified description of the propagation of thermal waves induced by a square-wave-modulated heating source. Generally, thermal waves are analyzed within the harmonic regime, where heat is delivered to the sample sinusoidally. In this scenario, the temperature distribution at the surface of a semi-infinite substrate ($\mathcal{W} \gg d$, where $\mathcal{W}$ is the width and $d$ is the thickness) is given by

\begin{equation}\label{eq:thermal_wave_harmo}
\Delta T(r,t)=\Re\left[{\frac{I_0}{4\pi r \kappa}e^{-r/\lambda}e^{i(\omega t-r/\lambda)}}\right],
\end{equation}

where $\Delta T(r,t)$ represents the temperature rise at a radial position $r$ and time $t$, $I_0$ is the amplitude of the power input, $\kappa$ is the thermal conductivity of the material, $\omega$ is the angular frequency of the excitation, and $\lambda=\sqrt{\frac{2D}{\omega}}$ is the thermal diffusion length, with $D=\frac{\kappa}{\rho c}$ the thermal diffusivity \cite{Carslaw_1959}.

\begin{figure}
    \centering
    \includegraphics[width=\linewidth]{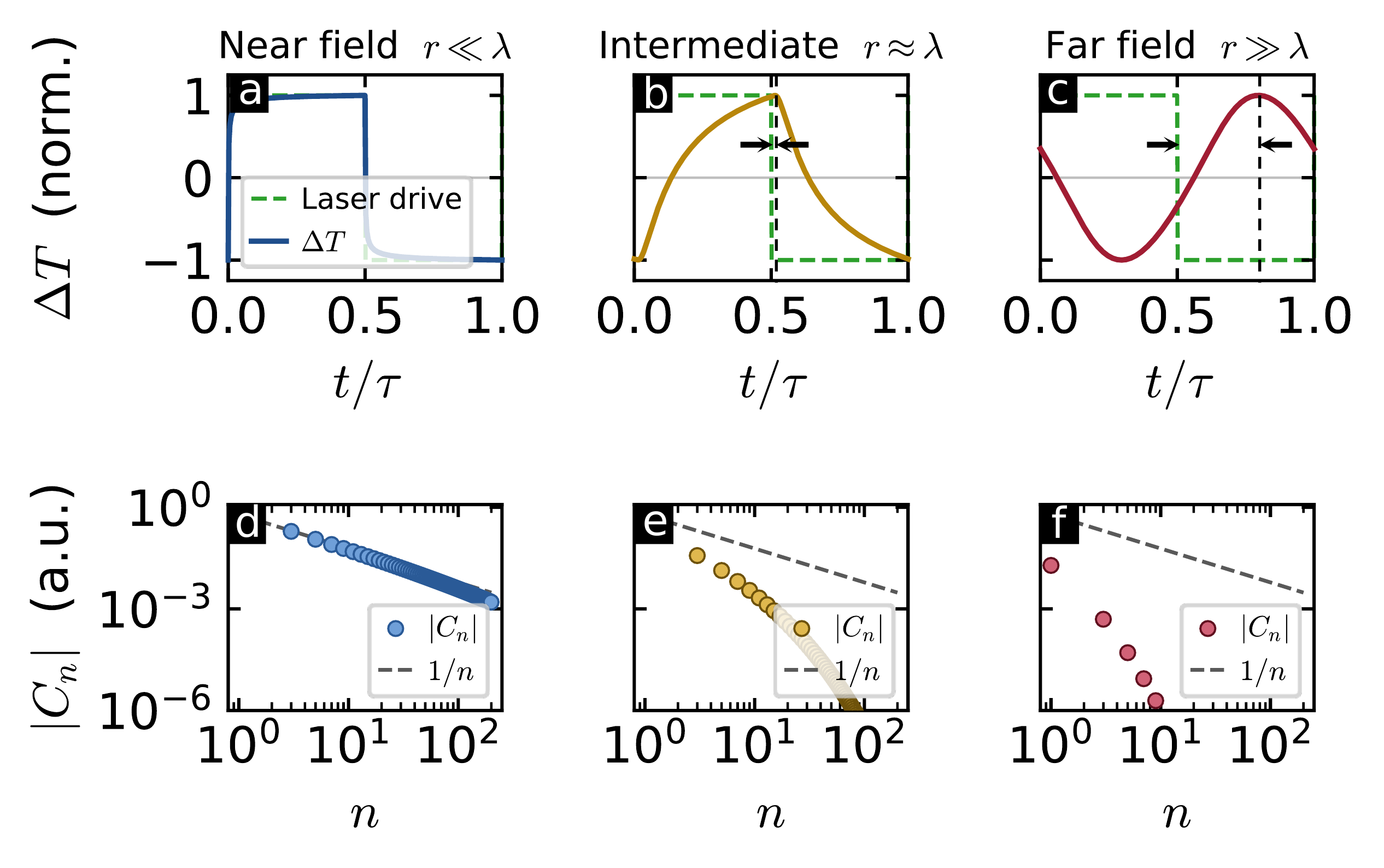}
    \caption{Temperature modulation induced by a chopped laser as a function of the reduced distance $r/\lambda$, computed from the Fourier expansion of Eq.~\eqref{eq:thermal_wave_harmo}. (a-c) Temperature response $\Delta T$ over one period of the excitation (solid line), together with the ideal square-wave laser drive (green dashed line). The delay between the switching moment and the maximum response is shown with arrows in (b,c). (d-f) Amplitude of the associated Fourier coefficients $|C_n|$ versus harmonic number $n$, compared with the $1/n$ decay of an ideal square wave (dashed line). (a,d) In the near-field regime ($r \ll \lambda$), the temperature closely retraces the square-wave drive and the coefficients follow the $1/n$ law up to high orders. (b,e) At intermediate distances ($r \approx \lambda$), the response acquires a slight phase delay relative to the drive and develops a saw-tooth shape, with a rapid relaxation after each transition followed by a gradual stabilization; the higher harmonics are progressively damped below the $1/n$ line. (c,f) In the far-field regime ($r \gg \lambda$), only the fundamental survives, giving a low-amplitude sinusoidal oscillation subject to a pronounced phase delay with respect to the excitation, and all higher harmonics are strongly suppressed.}
    \label{fig:thermal_waves}
\end{figure}

When employing a mechanical chopper, the heat input into the system is approximated by a square wave modulating the power between 0 and the laser power absorbed by the sample, $P_{0}$. Consequently, this anharmonic input can be expanded as a Fourier series of harmonic components:

\begin{equation}
I_{\text{abs}}(t)=\frac{I_0}{2}+\frac{2I_0}{\pi}\sum_{n=1,3,5,...}^{\infty}\frac{1}{n}\sin(n\omega t).
\end{equation}

By substituting this modulation into Eq.~\ref{eq:thermal_wave_harmo}, the resulting temperature response is expressed as

\begin{equation}
\begin{aligned}
\Delta T(r,t)&=\frac{I_0}{4\pi r\kappa}\left[ \frac{1}{2}\right.\\
&\left.+\frac{2}{\pi}\sum_{n=1,3,5,...}^{\infty}\frac{1}{n}e^{-r\sqrt{\frac{n\omega}{2D}}}\sin\left(n\omega t-r\sqrt{\frac{n\omega}{2D}}\right)\right].
\end{aligned}
\end{equation}

This expression corresponds to a low-pass filtered representation of a square wave, wherein the filter exponentially attenuates higher-order harmonics. Several distinct dynamic regimes emerge from this relation, which are illustrated in Fig.~\ref{fig:thermal_waves}. 

For $r \ll \sqrt{\frac{2D}{\omega}}$, the exponential term within the summation approaches unity, and the phase shift in the sinusoidal function becomes negligible for the dominant harmonics. Under these conditions, the summation reproduces the Fourier series of a square wave, causing the local temperature profile to closely track the square profile of the excitation source. This is apparent in Fig.~\ref{fig:thermal_waves}(a), where the temperature follows the drive almost instantaneously, and in Fig.~\ref{fig:thermal_waves}(d), whose Fourier amplitudes lie along the $1/n$ line up to high harmonic orders.

At intermediate distances, where $r \approx \sqrt{\frac{2D}{\omega}}$, the Fourier coefficients are attenuated by a factor of $e^{-\sqrt{n}}$. Concurrently, higher harmonics undergo distinct phase shifts relative to lower-order ones, resulting in a misalignment of their respective maxima. The superposition of these two effects introduces a pronounced asymmetry in the temporal response, yielding a saw-tooth waveform characterized by a rapid relaxation immediately following a transition in the source state, followed by a gradual stabilization. This behavior is shown in Fig.~\ref{fig:thermal_waves}(b): the response now lags the drive by a slight phase delay and loses its abrupt transitions, while the Fourier amplitudes in Fig.~\ref{fig:thermal_waves}(e) fall increasingly below the $1/n$ line as the exponential damping suppresses the higher harmonics.

Finally, in the asymptotic limit $r \gg \sqrt{\frac{2D}{\omega}}$, high-order harmonics are strongly damped, leaving only the fundamental component. In this regime, the temperature undergoes low-amplitude, purely sinusoidal oscillations. As illustrated in Fig.~\ref{fig:thermal_waves}(c), the surviving fundamental is subject to a large phase delay relative to the excitation, so that its maximum is markedly shifted in time with respect to the rising edge of the drive. Consistently, only the first few harmonics remain measurable in Fig.~\ref{fig:thermal_waves}(f), the rest having been extinguished by the exponential attenuation.

Notably, these anharmonic thermal waves offer, in theory, the distinct advantage of decoupling the respective contributions of $\kappa$ and $D$. In the near-field regime, the thermal diffusivity has a negligible influence on the local maximum temperature, which is predominantly governed by the prefactor inversely proportional to $\kappa$. This happens because the maximum temperature at the optical spot position results from a competition between heat generation (the laser power) and heat extraction (the thermal conductivity). Conversely, the diffusivity, together with the modulation frequency, sets the characteristic damping length of the oscillations. This spatial decay therefore determines the distance over which the thermal response transitions from a square wave to a saw-tooth profile, and asymptotically to a purely sinusoidal waveform.

\bibliography{references}

\end{document}